\documentclass[doublecol]{epl2}
\usepackage{graphicx}% Include figure files
\usepackage{bm}% bold math
\usepackage{amsmath} % \pmb bold special characters

\RequirePackage{color}
\definecolor{MyDarkGreen}{rgb}{0.02,0.60,0.06}
\def\naA{W}%\def\naA{{\mathsf W}}
\def\naE{{\mathsf E}}
\def\naB{{\mathsf B}}
\def\naphi{W_0}%\def\naphi{{\chi}}
\DeclareMathAlphabet{\mathssbf}{OT1}{cmss}{bx}{n}

\title{Spin superfluidity and spin-orbit gauge symmetry fixing}

\author{ Bertrand Berche\inst{1,2},  Ernesto Medina\inst{1,2} \and
	Alexander L\'opez\inst{2,3}}
\institute{
\inst{1} Statistical Physics Group, Institut Jean Lamour, Nancy Universit\'e,
UMR CNRS 7198\\ \phantom\quad 54506
Vand\oe uvre les Nancy, France\\
\inst{2} Centro de F\'\i sica, Instituto Venezolano de Investigaciones Cient\'\i ficas, Apartado 21874,\\ \phantom\quad Caracas 1020-A, Venezuela\\
\inst{3} Institute for Theoretical Physics, University of Regensburg, D-93040 
	Regensburg, Germany\\
}


\abstract{
The Hamiltonian describing 2D electron gas, in a spin-orbit active medium,
can be cast into a consistent non-Abelian gauge field theory leading to a
proper definition of the spin current. 
The generally advocated gauge symmetric version of the theory results in current densities 
that are gauge covariant, a fact that poses severe concerns on their physical nature. We 
show that in fact the problem demands gauge fixing, leaving no room to ambiguity in 
the definition of physical spin currents. Gauge fixing also allows for polarized edge 
excitations not present in the gauge symmetric case.
The scenario here is analogous to that of superconductivity gauge theory. 
We develop a variational formulation that accounts for the constraints 
between $U(1)$ physical fields and $SU(2)$ gauge fields and show that gauge 
fixing renders a physical matter and radiation currents and derive the particular 
consequences for the Rashba SO interaction.}

\begin{document}

\maketitle

\section{Introduction}
In condensed matter physics applications, the spin current is often considered
an ill-defined notion~\cite{Sonin10}, 
or at least a non conserved quantity  since angular momentum 
may be transferred via Spin-Orbit interactions (SOI) 
to the lattice. Obviously, the total angular 
momentum is conserved so it is illuminating to follow the route proposed by 
Yang and Mills in their description of the consequences of
the isotopic spin conservation\cite{YangMills54}: 
{\em The conservation of isotopic spin points to the existence of a
fundamental invariance law similar to the conservation of electric
charge. In the latter case, the electric charge serves as a source
of electromagnetic field. An important concept in this case is gauge
invariance which is closely connected with (1) the equation of
motion of the electromagnetic field, (2) the existence of a current
density, and (3) the possible interactions between a charged field
and the electromagnetic field.}
In the present context, we essentially start from point (3), 
since we build a consistent gauge theory from the SOI
(see Refs.~\cite{Goldhaber89,MineevVolovik92,FrohlichStuder93,RebeiHeinonen06,JinEtAl06,
LeursEtAl08,MedinaEtAl08,BercheEtAl09,Tokatly08,DartoraCabrera10}) 
and we deduce point (2), i.e. the definition of a conserved quantity, 
the dissipationless spin current density.
A problem then arises, due to the non-Abelian character of the underlying
gauge theory. The current density (color current in the context of high energy
physics, but here an angular momentum current density)
consists of two parts: a matter contribution plus the so-called radiation 
contribution, the existence of which is linked to the fact that the non-Abelian
gauge fields carry the corresponding color charge (this is also the 
origin of the
non linear character of the equations of motion satisfied by the gauge fields,
point (1) in Yang and Mills' quotation).
The matter contribution here corresponds to the spin current
and the radiation contribution is associated to the angular momentum carried
by the lattice degrees of freedom. 
The difficulty is that the conserved current density 
{\em has atrocious transformation properties
under a change of gauge} (Ramond~\cite{Ramond90}, explicitely 
given in Ref.~\cite{Hughes78}), and in the gauge symmetric formulation\cite{FrohlichStuder93,Tokatly08}, neither 
the total current density nor the two contributions separately (matter or radiation)  is gauge 
invariant. None of these densities is then 
acceptable as a physical quantity.

In this paper we draw attention to the fact that both the Pauli equation and 
the gauge formulations of the Rashba and 2D Dresselhaus Hamiltonian cannot be 
written correctly in a generally gauge invariant 
fashion\cite{LeursEtAl08,MedinaEtAl08}. We argue that this makes a very strong 
physical statement, since as discussed above, a gauge invariant Hamiltonian 
would not allow a locally gauge invariant spin current or spin polarization. We show that, in fact, the 
correct formulation contains and demands  gauge symmetry breaking that allows 
for non ambiguous %locally gauge invariant 
matter spin current. 
This is also related to the existence of a gauge potential  as a 
physical field in the same way as in superconductivity, required by the very 
identification made in \cite{Tokatly08} where it is directly related to a 
physical $U(1)$ electric field. Note that from a semantic point of view, 
we will follow Sonin's use of the term superfluidity, 
{\em superfluidity means only a possibility to transport
a physical quantity (mass, charge, spin, etc.) without 
dissipation~\cite{Sonin10}.}

\section{Yang-Mills formulation}

We first recall the essential steps of the construction of the gauge field
theory associated to SOI. Let us start by considering the Pauli Hamiltonian 
to order $v/c$ (in the International System of units)
\begin{eqnarray}
H&=&\frac{\left ({\mathbf p}-e{\mathbf A}\right)^2}{2m}+V
-\frac{e\hbar}{2m}{\boldsymbol \sigma}\cdot {\mathbf B} \nonumber \\
&&-\frac{e\hbar{\boldsymbol \sigma}\cdot{\bf E}\times 
\left ({\mathbf p}-e{\mathbf A}\right)}{4m^2c^2}
-\frac{ie\hbar^2{\boldsymbol \sigma}\cdot{\boldsymbol\nabla}\times {\bf E}}{8m^2c^2}
\mbox{} \nonumber \\
&&-\frac{e\hbar^2}{8m^2c^2}{\boldsymbol\nabla}\cdot{\bf E},\label{PauliFirst}
\end{eqnarray}
where the first and second terms are the kinetic energy, including the minimal
coupling to the electromagnetic field ($e=-|e|$ is the electron charge) 
and the substrate potential denoted 
by $V$, that can be assumed periodic. The third term is the Zeeman interaction
where $\mathbf B$ is the magnetic 
field  and $\boldsymbol \sigma$ is the vector of Pauli 
matrices ${\boldsymbol\sigma}=\sigma^a\hat {\bf u}_a$. The fourth and fifth 
terms are the spin-orbit interaction and 
the last one is the Darwin term (${\bf E}$ is the electric field and the 
bold font is used to denote vectors in ordinary space). 
 
In most of the applications considered, the rotor of the electric 
field is absent~\cite{CommentRotorE}, however, as it is discussed below, 
our main results would not depend on this assumption.   
In order to suggest an $SU(2)\times U(1)$ form we can 
rewrite the Hamiltonian,  following Jin, Li and Zhang\cite{JinEtAl06} as
\begin{eqnarray}
H&=&\frac{1}{2m}\left ({\mathbf p}-e{\mathbf A}-g{\bf W}^a\tau^a\right )^2
\nonumber \\
&&-\frac{g^2}{8m}{W}_i^a{W}_i^a
+ec{A}_0+gc{W}_0^a\tau^a.\label{PauliGauge}
\end{eqnarray}
One can identify the third term in the square as a new $SU(2)$ connection 
defined by $g{W}_i^a\tau^a=-(e\hbar/ 2mc^2)\epsilon_{aij}{ E}_j\tau^a$
(or ${W}_i^a=(e/ 2mc^2)\epsilon_{aji}{ E}_j$~\cite{Comment}), 
where the $\tau^a$ (half the Pauli matrices) 
are the symmetry generators for $SU(2)$ obeying the 
commutation relation $[\tau^a,\tau^b]=i\epsilon_{abc}\tau^c$, the symbol
$\epsilon_{abc}$ being the totally antisymmetric  tensor, and
$g=\hbar$. The relation between the spin 
operator and the corresponding generators is
$g\tau^a=s^a$  with spin one-half.     
The fourth term in equation~(\ref{PauliGauge}) represents the Zeeman 
energy and defines the time component of the $SU(2)$ connection,
$gc{W}_0^a\tau^a=-(e/m){\mathbf s}\cdot{\mathbf B}$ (or $W_0^a=-(e/mc)B^a$). 
Following the  conventional use for quadri-vectors we set
$W^a_\mu= (W_0^a,W_i^a)$  
and  $A_\mu= (A_0,A_i)$ with $A_0=V/ce$ in equation~(\ref{PauliFirst}). 
The Latin indices 
from the middle of the alphabet, $ijk$, denote the spatial coordinates and 
run over three values, Greek indices play the same role including also
$0$ as the time component, 
and the letters {\it a,b,c} denote the internal space 
indices for the generators. We always use the internal space indices as 
superindices, except in the antisymmetric tensor. 

This formulation differs crucially from that of 
ref.\cite{Sonin10,JinEtAl06,Tokatly08} %most importantly 
in the second term written
here as a function of the $SU(2)$ connection to evidence 
{\it gauge symmetry breaking} (GSB) in this Hamiltonian. The origin of this 
term is found in the fact that the SOI is linear in momentum.

The Lagrangian generating the correct Schr\"odinger equation is thus 
\begin{eqnarray}
{\mathcal L}&=& \frac{i\hbar}{2}\left (\psi^{\dagger}\dot{\psi}
-\dot{\psi}^{\dagger}\psi\right )
-\frac{\hbar^2}{2m}
\left [ {\mathcal D}_j\psi\right ]^{\dagger}\left[{\mathcal D}_j\psi\right]
\nonumber\\
&&-\psi^{\dagger}\left (\frac{-\hbar^2}{8m}{W}_i^a{W}_i^a+\hbar c{W}_0^a\tau^a
+e c{A}_0\right)\psi\nonumber\\
&&-\frac14{\varepsilon_0c^2}F_{\mu\nu}F^{\mu\nu}
-\frac14{\hbar c}{G}^a_{\mu\nu}{G}^{a\mu\nu},\label{eq-Lagrangian}
%\nonumber
\end{eqnarray}
where $\psi$ is a Pauli spinor, ${G}_{\mu\nu}^a 
=\partial_\mu{W}_\nu^a-\partial_\nu{W}_\mu^a
 -\epsilon_{abc}{W}_\nu^b{W}_\mu^c$ and ${F}_{\mu\nu} 
=\partial_\mu A_\nu-\partial_\nu A_\mu$ are the $SU(2)$ and $U(1)$ 
field tensors, respectively. 
The free non-Abelian field contribution 
$-\frac14{\hbar c}{G}^a_{\mu\nu}{ G}^{a\mu\nu}$ corresponds to the
spin-orbit correction to the energy in a rotationally invariant 
system~\cite{Tokatly08}.
%The new term $\frac{-g^2}{8m}{W}_i^b{W}_i^b$ 
%arises when building 
The covariant derivatives are defined by
${\mathcal D}_i=\partial_i-\frac{ie}{\hbar}A_i
-i {W}^a_i\tau^a$ for the space derivative and
${\mathcal D}_t=\partial_t+\frac{iec}{\hbar}A_0
+ic {W}^a_0\tau^a$ for the corresponding  time derivative. % that define the gauge potential. 
%For the Pauli Hamiltonian\cite{Tokatly08,MedinaLopezBerche}, 
%$g{W}_i^a\tau^a=-(e\hbar/2 mc^2)\epsilon_{iaj}{ E}_j\tau^a$,  
%where the $\tau^a$ are the symmetry generators for $SU(2)$. 
 {The first two lines of 
equation (\ref{eq-Lagrangian}) can be referred to as the matter part 
${\cal L}_{\rm mat.}$ (since they contain the spinor $\psi$)
and the last line as the pure field (or radiation) part 
${\cal L}_{\rm field}$. }

 {Let us illustrate our purpose with
the case of a quasi two-dimensional non interacting electron gas 
subject to both Rashba and Dresselhaus spin orbit interactions~\cite{Winkler}.
The components of the non-Abelian gauge field are in this case
\begin{eqnarray}
\frac{g}{2m}{\bf W}^a\tau^a=(\beta\tau^x-\alpha\tau^y)\hat{{\bf x}}
+(\alpha\tau^x-\beta\tau^y)\hat{{\bf y}}.\label{eqRashbaDresselhaus}
\end{eqnarray}
It is possible to measure and control the Rashba parameter $\alpha$
using gate voltages, e.g. in two dimensional GaAs/AlGaAs electron 
gas\cite{Nitta2,Shapers,Studer,MillerGoldhaberGordon},
$\alpha=b\langle E\rangle$, where $\langle E\rangle$ is 
the expectation value of the electric field at the 2DEG, and $b$ depends on 
the inverse of both the effective mass and the material gap\cite{Lommer}.  
The simple form for the Dresselhaus term (with parameter $\beta$) is obtained
with the assumption of strong confinement at the length scale $d$ 
of the electron gas, $k_F \ll \pi/d$\cite{Halperin} where $k_F$ is the Fermi 
wavector of the 
in plane electrons,  reducing 
cubic contributions to the dominant linear terms 
in equation (\ref{eqRashbaDresselhaus}).
%Let us also mention that in the case of constant SO parameters $\alpha$ and 
%$\beta$, the GSB term in equation (\ref{PauliGauge}) is a constant.

Under a global change of gauge, %equation~(\ref{eqWChange}) shows that
the gauge field (\ref{eqRashbaDresselhaus}) 
would rotate in spin space, i.e. the physical situation
would be different, with a SO interaction resulting from a rotated 
electric field.
Such a gauge transformation would change
the effective magnetic field around which the spin precesses,
${\bf k}\times{\bf E}$, and change the workings of SO based devices.
The very values of the SOI Rashba and Dresselhaus would be changed
one into another by such a transformation, while experimentally they have
definite values. This simple physical argument shows that a mechanism should
exist to prevent from such gauge transformations, 
and this is the subject of the 
next section.

}

\section{Gauge fixing}

It is clear that the Lagrangian density (\ref{eq-Lagrangian}) 
is not invariant under 
local $SU(2)-$gauge transformations $U=e^{i\tau^a\alpha^a(x)}$
\begin{eqnarray}
&&\psi \rightarrow U\psi,\\
%&&W_\mu \rightarrow U W_\mu U^{-1} -\frac{i\hbar}{g}(\partial_\mu U) U^{-1},
&&W_\mu^a\rightarrow W_\mu^a+\epsilon_{abc}W^b_\mu\alpha^c
	+\partial_\mu\alpha^a,\label{eqWChange}\\
&&G_{\mu\nu}^a\rightarrow G_{\mu\nu}^a+\epsilon_{abc}G^b_{\mu\nu}\alpha^c.
\end{eqnarray}
%as can be seen from 
The presence of the quadratic, in the gauge field, term 
restricts the gauge invariance to a smaller set of transformations. 
Let us focus the attention to this term,
\begin{eqnarray}
\frac{\hbar^2}{8m}\psi^\dag W^a_i W^{ai} \psi.
\end{eqnarray}
Under an infinitesimal local gauge transformation parameterised by 
$\alpha^a$, this contributes an 
additional term in the Langrangian,
%\begin{eqnarray}
$\frac{\hbar^2}{8m}\psi^\dag \left ( W^a_i \partial^i \alpha^a + (\partial_i \alpha^a) W^{ai} \right ) \psi + O(\alpha^2)$ 
%\end{eqnarray}
which can be written as a pure divergence,
%\begin{eqnarray*}
$\frac{\hbar^2}{4m}\partial_i\left(  W^{ai}\alpha^a \psi^\dag \psi \right)-\frac{\hbar^2}{4m}\left( (\partial_i W^{ai})\alpha^a \psi^\dag \psi +W^{ai}\alpha^a \partial_i \left(\psi^\dag \psi\right) \right)$
%\end{eqnarray*}
at the condition  that the gauge transformations are restricted to
\begin{eqnarray}
\partial_i W^{ai}=0 \hspace{0.5cm} \mbox{and} \hspace{0.5cm} \partial_i 
\left(\psi^\dag \psi\right)=0\label{GaugeCondns}.
\end{eqnarray}
The first condition is an $SU(2)$ version of the Coulomb gauge, leading here
to ${\boldsymbol\nabla}\times{\bf E}=0$ which is consistent with our choice of
simplification of the SO interaction. In a more general situation, one would
have $\partial_\mu W^{b\mu}=-(e/mc^2)[{\boldsymbol\nabla}\times{\bf E}+
\partial_t{\bf B}]_b=0$, also consistent with the Maxwell-Faraday 
law~\cite{LeursEtAl08}.
 The second constraint in equation~(\ref{GaugeCondns}) 
is the condition for fixing the norm of the wave function. It is worth 
noting that a spinor of the form $\psi=\begin{pmatrix}
0\\ \sqrt n e^{i\theta(x)}\end{pmatrix}$ satisfies such a conditions and leads
to a superfluid matter spin current 
(defined later) ${\bf J}^a=\frac{\hbar^2}{2m}
({\boldsymbol\nabla}\theta)n\delta^{a3}$. In the partition function
$Z=\int{\cal D}[\psi^\dagger]{\cal D}[\psi]e^{-\frac 1T\int d^3x\ \!{\cal L}}$,
this term contributes a high weight in the kinetic energy and in the mean
field approximation, the excitations with fluctuating phase can be
neglected, leaving zero bulk matter spin current, a situation
similar to superconductivity.
After the gauge transformation, the remaining pure divergence contributes 
to the action as a surface term, leaving the action covariantly 
transformed
%In the same spirit of the Chern-Simons theory, this lack of invariance can be 
%related to the gauge invariant form of the action, by rewriting the extra 
%terms as a choice of gauge plus a surface term.
\begin{eqnarray}
S&\rightarrow&\int d^4 x \ \!\mathcal{L}
+\frac{\hbar^2}{4m}\int dt \oint d\sigma_i   W^{ai}
\alpha^a \psi^\dag \psi \nonumber\\
&=&S+\hbox{surface terms}.
\end{eqnarray}
In a consistent gauge formulation, the surface term being gauge dependent 
should vanish, leading to boundary conditions at the edge of a finite system. 
This would imply that the non-Abelian gauge vector has no
component perpendicular to the surface, 
${\bf W}^a|_{{\rm edge}}\cdot\hat{\bf n}=0,\ 
\forall a$. This condition induces a similar constraint for the ``non-Abelian electric
field'' ${\mathssbf E}^a$ to be defined in the following section, and as
this field will be seen to obey Maxwell-like equations, the discontinuity
of its normal component across a boundary is associated to the existence
of a spin polarization sitting at the boundary. 
Vanishing normal components
thus do not allow for spin accumulation at the edges of a finite system, in
contradiction with experimental results~\cite{Awschalom}. 
From this argument, one has to 
conclude that there should be no gauge freedom, hence no surface contribution
to the action. This is obviously corroborated by the fact that the non-Abelian
gauge fields ${\bf W}^a$ and $W_0^a$ are defined in terms of the {\em physical}
$U(1)$ electric and magnetic fields.

The situation is similar to the case of superconductivity, for which there is 
also a mass term that generates, through 
gauge transformations, 
gauge-dependent bulk and surface contributions. Killing the bulk contribution
necessitates imposing the Coulomb gauge condition ${\boldsymbol\nabla}\cdot
{\bf A}=0$ and a condition on the superconducting wavefunction 
${\boldsymbol\nabla}|\varphi|^2=0$. There still remains a surface term which
disappears under  restricted gauge transformations at the condition 
that ${\bf A}$ be tangent to the boundary
of the system. This would then imply vanishing tangential components for
the magnetic field, hence no surface superconducting current. This is 
obviously a wrong conclusion which can only be discarded
if from the beginning one would not allow any gauge freedom. This
mechanism leads to the phase locking of the wave function, and renders the
gauge vector field physical, since it is automatically proportional to the
superfluid current.

Returning to the spin-orbit interaction, 
in order to obtain the equations of motion for the gauge fields, 
we have to take into account 
another peculiarity of the present problem, i.e. the fact that the $U(1)$ and
the $SU(2)$ gauge 
fields are not independent. The variational formulation has to include
constraints via Lagrange multipliers which allow to treat the different 
gauge fields as independent, 
%\begin{widetext}
\begin{eqnarray}
S_{\rm constr.}[A_\mu,W^a_\mu,\psi]=S[A_\mu,W^a_\mu,\psi]
\qquad\qquad\qquad\qquad&&\nonumber\\
\quad-\int dt\ \! d^3r\ \!\left[\mu^a(W_0^a+\frac e{mc} B^a)
+\lambda_i^a(W_i^a-\frac e{2mc^2}\epsilon_{aji}E_j)\right]&&\ \ \nonumber
\end{eqnarray}
%\end{widetext}
 where
$S[A_\mu,W^a_\mu,\psi]=
\int dt\ \! d^3r\ \!
{\cal L}(A_\mu,W^a_\mu,\psi)$.
%\begin{widetext}
%\begin{eqnarray}
%&&S_{\rm constr.}[\phi,A_k,\naphi^a,\naA^a_k]=S[\phi,A_k,\naphi^a,\naA^a_k]
%\nonumber\\
%&&\qquad\qquad-\int dt\ \! d^3r\left[\mu^b(\naphi^b-\frac em B^b)
%+\lambda_i^b(\naA_i^b-\frac e{2mc^2}\epsilon_{bji}E_j)\right],\\
%&&S[\phi,A_k,\naphi^a,\naA^a_k]=
%\int dt\ \! d^3r
%{\cal L}_{\rm tot.}(\phi,A_k,\naphi^a,\naA^a_k)
%\end{eqnarray}
%\end{widetext}
For the
usual electric and magnetic fields, we obtain the
following equations of motion (Maxwell equations) via variation w.r.t the 
Abelian gauge field components:
\begin{eqnarray}
&&-\rho_{\rm mat.}+\varepsilon_0\partial_k E_k+\frac e{2mc^2}\partial_k
\epsilon_{aki}\lambda^a_i=0,\label{eqMaxwell1}\\ 
&&J_i-\varepsilon_0 c^2\epsilon_{ijk}\partial_j B_k
+\varepsilon_0\partial_t E_i  
-\frac e{mc}\epsilon_{aik}\partial_k\mu^a\nonumber\\
&&\qquad\qquad\qquad\qquad+\frac e{2mc^2}\partial_t\epsilon_{aik}
\lambda^a_k=0,\label{eqMaxwell2}
\end{eqnarray}
with $\rho_{\rm mat.}=e\psi^\dagger\psi$ and $J_i=\frac{-ie\hbar}{2m}
(\psi^\dagger(\partial_i\psi)-(\partial_i\psi^\dagger)\psi)
-\frac{e^2}m A_i\psi^\dagger\psi$.
In equation~(\ref{eqMaxwell1}), 
we can define a polarisation in terms of the
Lagrange multipliers, 
$P_k=\frac e{mc^2}\epsilon_{aki}\lambda^a_i$. This dielectric
polarization is associated to bound charges and
there appears then a 
contribution $-\partial_k P_k$ to the total charge density. This polarization 
also 
contributes the second equation of motion~(\ref{eqMaxwell2}) 
through a term added to the
ordinary current density, $\partial_t P_i$, as well as an additional term
$\epsilon_{ika}\partial_k m^a$, with $m^a=-\frac e{mc} \mu^a$, describing
``Amp\`erian'' currents.
The electric charge and current density are thus
modified by the SO$+$Zeeman terms and the usual Maxwell 
equations become~\cite{WangXiaSuMa}:
%\begin{equation}
%\begin{array}{l}
${\boldsymbol\nabla}\cdot{\bf E}=\frac{\rho_{\rm mat.}}
{\varepsilon_0}-{\boldsymbol\nabla}\cdot{\bf P}$, 
%\\
and
${\boldsymbol\nabla}\times{\bf B}
-\varepsilon_0\mu_0\partial_t{\bf E}
=\mu_0[{\bf J}+\partial_t{\bf P}+{\boldsymbol\nabla}\times{\bf m}]$.
%\end{array}
%\end{equation}

The variation with respect to the non-Abelian gauge fields leads to 
\begin{eqnarray}
%0&=&\frac{\delta S_{\rm constr.}}{\delta\naphi^a}%&=&\nonumber\\&=&
&&-\rho_{\rm mat.}^a+\frac\hbar{2c}\epsilon_{abc}\naA^b_k\naE^c_k
-\frac\hbar c\partial_k\naE^a_k+\mu^a=0,\\
%0&=&\frac{\delta S_{\rm constr.}}{\delta\naA^a_i}%&=&\nonumber\\&=&
&&J_i^a-\frac\hbar{2c}\epsilon_{abc}[\naphi^b\naE^c_i+c^2\epsilon_{ijk}
\naA^b_j\naB^c_k]\nonumber\\
&&\qquad\qquad\quad-\hbar c\epsilon_{ijk}\partial_j\naB^a_k+\frac\hbar c\partial_t\naE_i^a
-\lambda^a_i=0,
\end{eqnarray}
where instead of the field strength tensor $G^a_{\mu\nu}$, we have defined
the non-Abelian ``electric'' and ``magnetic'' fields (written in sanserif)
\begin{equation}
\begin{array}{l}
{\mathssbf E}^a=-\partial_t{\bf W}^a-c{\boldsymbol\nabla}\naphi^a+\frac 12c\epsilon_{abc}
\naphi^b{\bf W}^c,\\
{\mathssbf B}^a={\boldsymbol\nabla}\times{\bf W}^a
+\frac 12\epsilon_{abc}{\bf W}^b\times{\bf W}^c.
\end{array}
\end{equation}

Note that we have defined the matter spin density and superfluid 
spin current as
\begin{eqnarray}
\rho_{\rm mat.}^a&=&
% {-\frac 1c\frac{\partial{\cal L}_{\rm mat.}}{\partial W^a_0}=}
\psi^\dagger\hbar\tau^a\psi,\label{eq-rhomat} \\
J_i^a&=&
% {\frac{\partial{\cal L}_{\rm mat.}}{\partial W^a_i}=}
-\frac{i\hbar^2}{2m}(\psi^\dagger\tau^a(\partial_i\psi)-
(\partial_i\psi^\dagger)\tau^a\psi),%\qquad
\label{eq-jmat}
\end{eqnarray} 
where, as already discussed in 
ref.~\cite{MedinaEtAl08}, the presence of the gauge symmetry breaking term
in the Lagrangian density exactly compensates the ``diacolor'' contribution to
the matter current and renders zero spin conductivity\cite{Inoue,Raimondi,Dimitrova}, as expected
from one electron Hamiltonians with linear SOI.
We also recover the appearence of the ``radiation'' spin density 
and the ``radiation'' spin current density which would follow from the 
Noether theorem,
\begin{eqnarray}
\rho_{\rm rad.}^a&=&% {-\frac 1c\frac{\partial{\cal L}_{\rm field}}{\partial W^a_0}=}
-\frac\hbar{2c}\epsilon_{abc}\naA^b_k\naE^c_k,
\label{eq-rhorad}\\ 
{\cal J}^a_i&=&% {\frac{\partial{\cal L}_{\rm field}}{\partial W^a_i}=}
-\frac\hbar{2}\epsilon_{abc}(\naphi^b\naE^c_i+c\epsilon_{ijk}
\naA^b_j\naB^c_k),%\qquad
\label{eq-jrad}
\end{eqnarray} 
in terms of which the
corresponding Yang-Mills-Maxwell equations become
\begin{eqnarray}
%\begin{array}{l}
&&{\boldsymbol\nabla}\cdot{\mathssbf E}^a=\frac c\hbar(\rho^a_{\rm mat.}+\rho^a_{\rm rad.}
-\mu^a),\label{eqYMM1}\\
&&{\boldsymbol\nabla}\times{\mathssbf B}^a
-\frac 1{c^{2}}\partial_t{\mathssbf E}^a
=\frac{1}{\hbar c}({\bf J}^a+{\boldsymbol{\cal J}}^a-{\boldsymbol\lambda}^a),
\label{eqYMM2}
\end{eqnarray}
where the calligraphic characters denote 
the radiation contributions while roman fonts are used for the matter
quantities.
The occurrence of the radiation contribution to the conserved Noether current 
are directly linked to the existence of the SO and Zeeman interactions and 
these terms are mandatory in order to satisfy the conservation of the total 
angular momentum.
The matter contributions to this conserved current is the usual spin current
associated to the free electrons, while the radiation part corresponds to the 
angular momentum carried by the lattice.
%The non-Abelian equations without sources take the form
%\begin{eqnarray}
%&&{\boldsymbol\nabla}\times\vec\naE^a+\partial_t\vec\naB^a=
%\frac 12\epsilon_{abc}(\partial_t(\vec\naA^b\times\vec\naA^c)
%+c{\boldsymbol\nabla}\times (\naphi^b\vec\naA^c)),\label{ym3}\nonumber\\
%&&{\boldsymbol\nabla}\cdot\vec\naB^a=
%\frac 12\epsilon_{abc}{\boldsymbol\nabla}\cdot(\vec\naA^b\times\vec\naA^c)
%.\nonumber 
%\end{eqnarray}

 {
In order to illustrate how the above contributions are at 
work in a simple case,
let us again contemplate the case of a 2D electron 
gas with only uniform Rashba SO interaction, immersed in a perpendicular
magnetic field. Using the notations $\kappa=2m\alpha/\hbar$ and $\Omega=eB/m$,
 we write the components of the non-Abelian gauge field 
 (\ref{eqRashbaDresselhaus}) as 
${\bf W}^1=-\kappa\hat {\bf y}$,
${\bf W}^2=\kappa\hat {\bf x}$ and $W_0^3=\Omega/c$.
The components of the non-Abelian electric and magnetic fields follow, and
allow to compute the non vanishing components of the radiation spin 
density $\rho^3_{\rm rad.}=-\frac \hbar{2c}
\Omega\kappa$ and spin current ${\boldsymbol{\cal J}}^1_{\rm rad.}=
-\frac\hbar{2c}\kappa(\frac 12\Omega^2-\kappa^2c^2)\hat {\bf y}$ and
${\boldsymbol{\cal J}}^2_{\rm rad.}=
\frac\hbar{2c}\kappa(\frac 12\Omega^2-\kappa^2c^2)\hat {\bf x}$. The
radiation spin current induces an electric polarization given by 
\begin{equation}
{\boldsymbol{\cal P}}=\frac{e\hbar}{2mc^3}\kappa(\frac 12\Omega^2-\kappa^2c^2)
\hat{\bf z}.\label{eqPolarization}
\end{equation} 
Note that an expression similar to 
equation~(\ref{eqPolarization}) was obtained in Ref.~\cite{WangXiaSuMa}.
The spin polarization generates a magnetic moment whose
radiation contribution is given by 
\begin{equation}
{\boldsymbol{\cal M}}=\frac {e\hbar}{2mc}
\Omega\kappa^2\hat{\bf z}.
\end{equation} The interesting result here is that a pure
Zeeman term would only produce a spin precession, hence would not polarize
the electrons, but in association with SO interaction which acts with a 
perpendicular torque (relaxation of the spin polarization along the
precession axis  
was first written phenomenologically
by Landau and Lifshitz~\cite{LL,LandauLifshitz}), 
the spin orientation relaxes and a net polarization 
occurs.

The electric polarization and the magnetic moment are physical observable
quantities, and a gauge transformation, if contemplated, 
would change the orientation
of these fields, and hence the physical situation. 
%This is also an illustration of the necessity 
%for keeping a fixed SO gauge.
}

 {From more general arguments, }the necessity of a fixed gauge is 
clear when one notices that neither the
matter (\ref{eq-rhomat},\ref{eq-jmat}) nor radiation 
(\ref{eq-rhorad},\ref{eq-jrad}) 
spin polarizations or currents are gauge invariant. In particular we have
covariance of the matter current
$J^a_i\longrightarrow J^a_i-\epsilon_{abc}\alpha^b J^c_i$ and an even more 
complicated transformation law for the radiation current~\cite {Hughes78}.
Concerning the radiation contribution, this is a general feature of non-Abelian gauge 
fields~\cite{Ramond90,Hughes78}. The case of the matter spin current
is more perplexing and the present situation is due to the cancellation of
the diacolor contribution. In the $U(1)$ case, both the paramagnetic and
the diamagnetic current densities are gauge dependent, but their sum is not. 
In the case of superconductivity, where only one of the contributions survives
(the diamagnetic one), the current is made physical by the fact that the gauge is 
fixed~\cite{Weinberg}. We have in a sense a parallel situation here, where again, gauge fixing
is required.

Now the role of the Lagrange multipliers needs to be analyzed more completely.
From the Yang-Mills-Maxwell equations, one can form a continuity equation
describing the conservation of the total angular momentum density. Taking the
divergence of Eq.~(\ref{eqYMM2}), and using
Eq.~(\ref{eqYMM1}), one gets
\begin{equation}\partial_t(\rho^a_{\rm mat.}+\rho^a_{\rm rad.})+{\boldsymbol\nabla}\cdot
({\bf J}^a+{\boldsymbol{\cal J}}^a)=\partial_t\mu^a+{\boldsymbol\nabla}\cdot
{\boldsymbol\lambda}^a.\end{equation}
The Lagrange multipliers formally appear in the continuity equation, but 
physically, the l.h.s. already contains all possible sources of angular 
momentum in the problem, i.e.  {i)} matter spin density (or all free 
electrons contributions from the conduction band encoded in the spinor $\psi$),
and  {ii)} radiation contributions (or here angular momentum 
transfered to the lattice via SO interaction). 
 {The firts contribution accounts for conducting electrons 
contribution, since these latter electrons (e.g. from an $s$-band) do not 
contribute through an orbital angular momentum while the second contribution  
includes all external fields terms from the Lagrangian density 
(\ref{eq-Lagrangian}). In the general case, there could also exist an angular 
momentum density associated to the ordinary $U(1)$ fields~\cite{Rohrlich},
but since we did not incorporate
any couplings of these fields to matter to allow for transfer of angular 
momentum, they should not enter the continuity equation.
}

There could also exist angular momentum density
associated to the ordinary $U(1)$ fields~\cite{Rohrlich}, 
but since we did not incorporate
any couplings of these to matter to allow for transfer of angular 
momentum, they should not enter the continuity equation. So eventually the
conservation of angular momentum in the system reads as
\begin{equation}
\partial_t(\rho^a_{\rm mat.}+\rho^a_{\rm rad.})+{\boldsymbol\nabla}\cdot
({\bf J}^a+{\boldsymbol{\cal J}}^a)=0.\label{eqnConservation}\end{equation}
This equation is often discussed in the literature in the form
$\partial_t\rho^a_{\rm mat.}+{\boldsymbol\nabla}\cdot
{\bf J}^a={\it torque\ density}$, considering that we can arbitrarily
modify the content of the matter spin current and that of the torque 
density in such a way that the equation is always satisfied. Our claim here
is that such a freedom does not exist, and we can indeed interpret the
radiation contributions, when written at the r.h.s of  equation
(\ref{eqnConservation}), as a 
torque density, but the respective definitions of matter and radiation
contributions are fixed.

 {
\section{Physical discussion and measurements}
A key point of our argument in this letter is the fact that due to the 
$SU(2)$ symmetry of the gauge group for the SO interaction, 
gauge transformations 
if performed, would {\emph change measurable physical quantities} (as a 
result, these transformations are of course not allowed). It is thus 
interesting to discuss in more detail possible measurements of such 
quantities. 

One of the primary quantities touched by such gauge transformations would be 
the spin polarization and spin current density. 
The electric polarization induced by the presence of spin current is an 
interesting feature for the detection of 
spin currents. In the same vein, the detection of charge imbalance was  
proposed a few 
years ago by Valenzuela and Tinkham~\cite{ValenzuelaTinkham}, when they
observed that due to the Onsager reciprocal relations between 
spin and charge currents, a   measurable  charge
asymmetry is expected in the presence of a spin current.
Let us also mention that the cubic dependence on the Rashba parameter of the 
dielectric polarization (\ref{eqPolarization}) 
induced by the radiation current should help in 
discriminating between matter and radiation contributions to the spin current 
density.

An even more transparent example is given by the famous experiments of the
beginning of the XXth century based on gyromagnetic 
phenomena~\cite{Barnett1935,Frenkel79}, and among them the
rotation of the sample in the Einstein-de Haas experiment. As we have argued, 
the continuity equation can be re-interpreted in terms of a torque density 
when all radiation contributions are shifted on the r.h.s. of the 
conservation equation. This torque density resulting from a re-writing of 
the radiation polarization and current (in the case of the Rashba material 
treated as an example in this letter it would just be 
$-\partial_t\rho^a_{\rm rad.}=\frac\hbar{2c}\kappa\dot\Omega$), it would 
thus be affected by a tentative gauge transformation. The ``reality'' of this 
torque density is demonstrated  in the Einstein-de Haas 
experiment,
 where a magnetic field applied to an initially non magnetized ferromagnet 
induces a spin polarization, hence a spin angular momentum then transferred 
to the lattice to ensure conservation of the total angular momentum. As a 
result, one obtains an orbital motion of the lattice. 
}

A very illuminating mechanical analog of angular
momentum currents, relevant to the previous argument, is 
discussed in reference \cite{Sonin10}: If one considers a solid cylinder made
of any particular material and twists it by a net angle $\theta$ at one end, we
generate a torsion. If one then attaches the two ends of the cylinder, thereby making a doughnut, the resulting
torsion will be built into the material and cannot be relaxed except by plastic deformations.
This is an example of a persistent angular momentum current and a
source torque proportional to the gradient of the twisting angle. In the sense
of the discussion before, once the material has been chosen and the
end to end twisting angle is fixed then one has no freedom in changing
the torque or the matter angular momentum current. This a clear example
of a mechanical fixed gauge situation.

\section{Conclusions}
As we see, the variational definition of the currents in terms of derivatives of the Lagrangian 
is an unambiguous way to define a current, since its canonical 
form does not depend on an ad hoc proposition. Nevertheless, the variational principle by 
itself is not enough to solve the ambiguity of the spin current definition.
To be consistent, we need also a symmetry broken formulation (or fixed gauge),
 otherwise, gauge transformations would change the matter (spin) and radiation
 (torque) content i.e. the matter currents are gauge covariant. A similar 
gauge symmetric scenario to the one presented here is materialized in 
superfluid condensates of neutral Bosons in the Helium 3 B phase, where one 
encounters a $SU(2)$  Anderson-Higgs mechanism\cite{FrohlichStuder93}. 
In reference\cite{Tokatly08}, Tokatly argues on the fact that the variational 
definition of the current density leaves no room for any ambiguity on the 
definition of the spin current density. We would like to stress that this 
statement is only valid in the formulation discussed here, in a 
gauge symmetry broken formulation.


\begin{thebibliography}{}
\bibitem{CommentRotorE}It is important to note here that even in the case of
	a non uniform electric field, (or more precisely a non vanishing
	${\boldsymbol\nabla}\times{\bf E}$), the gauge formulation given here
	is valid, with the same definition of the non-Abelian gauge fields.
	This is an advantage of the formulation on the direct Hamiltonian
	formalism where the rotor of ${\bf E}$ also implied for the Rashba parameter
	${\boldsymbol\alpha}$ is often neglected.
\end{thebibliography}
\end{document}